\documentclass[3p,final,times]{elsarticle}

\usepackage{amssymb}
\usepackage{amsmath}
\usepackage{bbm}
\usepackage{url}
\usepackage{lineno}

\journal{arXiv.org}

\usepackage{numcompress}\biboptions{authoryear}

\begin{document}

\begin{frontmatter}

\title{Computing the aggregate loss distribution based on numerical inversion of the compound empirical characteristic function of frequency and severity} 

\author[um]{Viktor Witkovsk{\'y}\corref{cor1}}
\ead{witkovsky@savba.sk}
\cortext[cor1]{Corresponding author. Tel.: +421 2 59104530; Fax: +421 2 54775943.}
\address[um]{Institute of Measurement Science, Slovak Academy of Sciences, D{\'u}bravsk{\'a} cesta 9, 841\,04 Bratislava, Slovakia}

\author[bratislava,bystrica]{Gejza Wimmer}
\address[bratislava]{Mathematical Institute, Slovak Academy of Sciences, Bratislava, Slovakia}
\address[bystrica]{Faculty of Natural Sciences, Matej Bel University, Bansk\'a Bystrica, Slovakia}
\ead{wimmer@mat.savba.sk}

\author[uk]{Tomas Duby}
\address[uk]{OAA Computing, Bicester, Oxfordshire, United Kingdom}
\ead{tomy@oaacomputing.co.uk}


\begin{abstract}
A non-parametric method for evaluation of the aggregate loss distribution (ALD) by combining and numerically inverting the empirical characteristic functions (CFs) is presented and illustrated. This approach to evaluate ALD is based on purely non-parametric considerations, i.e., based on the empirical CFs of frequency and severity of the claims in the actuarial risk applications. This approach can be, however, naturally generalized to a more complex semi-parametric modeling approach, e.g., by incorporating the generalized Pareto distribution fit of the severity distribution heavy tails, and/or by considering the weighted mixture of the parametric CFs (used to model the expert knowledge) and the empirical CFs (used to incorporate the knowledge based on the historical data --- internal and/or external). Here we present a simple and yet efficient method and algorithms for numerical inversion of the CF, suitable for evaluation of the ALDs and the associated measures of interest important for applications, as, e.g., the value at risk (VaR). The presented approach is based on combination of the Gil-Pelaez inversion formulae for deriving the probability distribution (PDF and CDF) from the compound (empirical) CF and the trapezoidal rule used for numerical integration. The applicability of the suggested approach is illustrated by analysis of a well know insurance dataset, the Danish fire loss data.
\end{abstract}


\end{frontmatter}


\section{Introduction}

In financial risk management, estimation of the operational risk capital under the loss distribution approach requires evaluation of the \emph{aggregate loss distribution} (ALD). Similarly,  the collective risk models (CRM) in insurance require evaluation of insurance portfolio ALD in a certain period of time, defined as a compound distribution of the intensity of the claims (frequency) and their sizes (severity). For more details see, e.g., \cite{Hogg1984}, \cite{embrechts2013modelling}, \cite{kaas2008modern}, and also \cite{schmidliaccumulated}, \cite{rolski2009stochastic}, \cite{Roncalli2016}.

An important measure of interest in such cases is the value at risk (VaR) which is typically defined as the $99.9\%$ quantile of the ALD. Frequently, the exponential, gamma, log-normal, log-logistic or Pareto distributions are used as the continuous severity distributions, and the Poisson, binomial, or negative binomial distributions are used as the discrete frequency distributions, with their respective parameters fitted from the available observed data and/or derived based on the expert knowledge/judgment.

The \emph{aggregate loss} in collective risk models, say, 
\begin{equation}\label{eq01}
	S = \sum_{j=1}^N X_j,
\end{equation}
is modeled as a sum of stochastic number $N$ of identically independently distributed (i.i.d.) random variables (RVs) $X_1,\dots,X_N$ which represent the claim sizes further modeled by $X_j \sim F_X$. By $F_X$ we denote the probability distribution, in particular the cumulative distribution function (CDF) of a continuous severity distribution which is independent of $N\sim F_N$, the random number of insurance claims generated in the given time period, and $F_N $ denotes the probability distribution of a discrete frequency distribution. Obviously, the aggregate loss $S = 0$ if $N = 0$.

The probability distribution of the aggregate loss (\ref{eq01}) is then a mixture distribution,
\begin{equation}\label{eq02}
	F_S = \sum_{n=0}^\infty \Pr(N=n) F_X^n, 
\end{equation}
where $\Pr(N=n)$ denotes the probability of the random event that $N=n$, and $F_X^n$ denotes the $n$-times convolved distribution function $F_X$ (with $F_X^0$ being the degenerate Dirac distribution concentrated at $0$, by definition). 

In general, assuming portfolio with $p$ independent event type and/or business line cells (with their aggregate losses $S_i$, $i = 1,\dots,p$), we define the \emph{aggregate loss of the (compound) portfolio}, say $L$, as
\begin{equation}\label{eq03}
	L = \sum_{i=1}^p S_i, 
\end{equation}
where $S_i= \sum_{j=1}^{N_i} X_{i,j}$ are mutually independent, with independent claim frequencies, $N_i\sim F_{N_i}$, and claim severities, $X_{i,j} \sim F_{X_i}$, for all $i = 1,\dots,p$ and $j = 1,\dots,N_i$.

The requirement on mutual independence of the involved RVs can be relaxed and naturally generalized for situations where the claim frequencies $N_i\sim F_{N_i}$, $i = 1,\dots,p$, are correlated and their joint distribution is given. In particular, \cite{ambagaspitiya1998distribution} considered a family of discrete multivariate distributions, where the $p$-variate discrete random vector $N = (N_1,\dots,N_p)^T$ can be represented by $N = A M$, and where $A$ is a $p \times k$ non-negative integer element matrix and $M = (M_1,\dots,M_k)^T$ is a $k \times 1$ column vector whose components RVs are independent. The situation with combining the correlated RVs is, however, not discussed in more details in this paper.

The closed-form expression of the CDF, such as defined in (\ref{eq02}), is typically not available for the aggregate loss $S$ or $L$ in (\ref{eq01}) and (\ref{eq03}). Thus, evaluation of these distributions relies on numerical methods. See \cite{Shevchenko2010} for an overview of available numerical algorithms that can be successfully used to calculate the ALD, including the Monte Carlo, Panjer recursion and Fourier transformation methods. \cite{heckman1983calculation} described a specific (mathematically convenient) model and method that numerically inverts the characteristic function (CF) of an ALD. 

In general, methods for approximate numerical inversion of CF, based on discrete Fourier transform and the fast Fourier transform (FFT) algorithm, can be used alternatively, see, e.g., \cite{feng2013inverting} or \cite{Witkovsky2016}. For further details on the FFT-based approach see the results in \cite{Huerlimann2013}. For methods based on fractional fast Fourier transform (FRFT), see \cite{Bailey1991}, \cite{Carr1999}, \cite{Chourdakis2004}, \cite{Held2014}, \cite{Kim2010}. 

Here we shall discuss in more details the methods and the algorithms for evaluating the required probability density function (PDF) and the CDF by numerical inversion of the associated CF using a trapezoidal quadrature rule, which is sufficiently precise for most practical situations, as well as their MATLAB implementation, the \emph{characteristic functions toolbox} (\texttt{CF Toolbox}).

Numerically more accurate {MATLAB} algorithms for inversion of the CFs used in specific CRM with selected parametric frequency and severity distributions are suggested and implemented elsewhere, see \cite{Duby2017}.

In this paper, we focus primarily on non-parametric considerations, i.e., on models based on combining the empirical CFs of claims frequency and severity in the actuarial risk applications, and their numerical inversion for evaluating the CDF and/or VaR. 

The basic non-parametric methods can be naturally generalized to a more complex semi-parametric modeling approach. For example, by combining the parametric frequency distributions with the empirical severity distributions, and/or by incorporating the generalized Pareto distribution fit of the severity distribution heavy tails. Or, by considering the weighted mixture of the parametric CFs (this is used to model and incorporate the expert knowledge) and the empirical CFs (used to include the knowledge based on internal and/or external historical data). In fact, the components of the compound ALD CFs can be represented either by parametric or empirical CFs and/or by their weighted mixtures.

As we shall argue and illustrate by analysis of a real data example in Section~\ref{sec04}, the suggested approach based on empirical distributions in combination with  parametric models used to model the heavy tails of the severity distributions and/or to include expert knowledge, gives better modeling flexibility than the standard parametric modeling approach. Moreover, evaluating the ALD by numerical inversion of the associated CF is computationally more efficient in comparison with other numerical methods, as, e.g., the  Monte Carlo methods or the Panjer recursion.

The rest of the paper is organized as follows: In Section~\ref{sec02} we shall present methods for computing and combining the characteristic functions used for modeling the ALDs. In Section~\ref{sec03} we introduce the Gil-Pelaez inversion and its implementation based on using the trapezoidal rule. Their applicability is illustrated in Section~\ref{sec04}, where the ADLs and VaRs are computed for real data. Discussion and concluding remarks are in Section~\ref{sec05}.

\section{The ALD characteristic functions}\label{sec02}

\subsection{Parametric CF}

For a scalar RV $X$ the characteristic function is defined as the expectation value of the transformed RV $\mathrm{e}^{\mathtt{i}tX}$, i.e. 
\begin{equation}\label{eq04}
\mathop{\mathrm{cf}}\nolimits_{X}(t) = \mathop{E}\left(\mathrm{e}^{\mathtt{i}tX}\right),  
\end{equation}
where $\mathtt{i}$ denotes the imaginary unit defined by $\mathtt{i} = \sqrt{-1}$, and $t \in \mathbb{R}$ is the argument of the CF. 

In particular, CF of a continuous univariate RV $X$, $\mathop{\mathrm{cf}}\nolimits_{X}(t)$, with its probability distribution $F_X$ (i.e.~$X\sim F_X$) and its probability density function $\mathop{\mathrm{pdf}}\nolimits_{X}(x)$, is given as the (inverse) Fourier transform of its PDF, 
\begin{equation}\label{eq05}
	\mathop{\mathrm{cf}}\nolimits_{X}(t) = \int_{-\infty}^\infty \mathrm{e}^{\mathtt{i}tx} \mathop{\mathrm{pdf}}\nolimits_{X}(x)\,dx, \ \ t \in \mathbb{R}.
\end{equation}
Note that since PDF is a real function, the CF is Hermitian, i.e.~$\mathop{\mathrm{cf}}\nolimits_{X}(-t) = \overline{\mathop{\mathrm{cf}}\nolimits_{X}}(t)$.

Analytical expressions of the CF are known for many standard probability distributions, see e.g.~\cite{lukacs1970characteristics}, or can be derived by using a suitable computer algebra system, as e.g.~{MATHEMATICA}. On the other hand, if the analytical form of the CF is unknown or it is too complicated, as it depends on nonstandard special functions and/or complicated series expansions (as is the case for the log-normal, log-logistic and Pareto distributions), such CFs can be still evaluated numerically, either directly from its definition (\ref{eq04}) or (\ref{eq05}), or other suitable alternative representation. 

For example, by using the half-space Fourier integral transformation for a positive continuous random variable $X$ (i.e. with $X\geq 0$) with its PDF given by an analytical function $\mathop{\mathrm{pdf}}\nolimits_{X}(z)$, which is well defined for complex $z\in \mathbb{C}$ and decays at infinity, 
we get
\begin{equation}\label{eq06}
	\mathop{\mathrm{cf}}\nolimits_{X}(t) = \int_{0}^\infty \frac{\mathtt{i}}{t} \mathop{\mathrm{pdf}}\nolimits_{X}\left(\frac{\mathtt{i}x}{t}\right)\mathrm{e}^{-x}  \,dx, \ \ t \in \mathbb{R},
\end{equation}
see, e.g., \cite{asheim2013complex}. Moreover, by using a suitable stabilizing transformation from $(0,\infty)$ to $(0,1)$, the CF can be numerically evaluated at arbitrary $t \in \mathbb{R}$ by using  any simple (Gaussian) quadrature rule of a well behaved integral,
\begin{equation}\label{eq07}
	\mathop{\mathrm{cf}}\nolimits_{X}(t) = \int_{0}^1 \frac{\mathtt{i}}{t} \mathop{\mathrm{pdf}}\nolimits_{X}\left(\frac{\mathtt{i}}{t}\left(\frac{x}{1-x}\right)^2\right) \frac{2x \mathrm{e}^{-\left(\frac{x}{1-x}\right)^2}}{(1-x)^3}  \,dx.
\end{equation}
This method can be used to evaluate CF of several continuous distributions listed in Table~\ref{tab1}.

In general, the CF of a weighted sum of independent random variables is calculated in the following way: 

Let $Y = c_1 X_1 + \cdots + c_n X_n$ be a weighted sum of independent RVs $X_1, \dots, X_n$, with $c_1, \dots, c_n$ known positive constants for fixed $n$, and known CFs $\mathop{\mathrm{cf}}\nolimits_{X_1}(t),\dots,\mathop{\mathrm{cf}}\nolimits_{X_n}(t)$. Then the CF of RV $Y$, $\mathop{\mathrm{cf}}\nolimits_{Y}(t)$, is given by  
\begin{equation}\label{eq08}
\mathop{\mathrm{cf}}\nolimits_{Y}(t) = \mathop{\mathrm{cf}}\nolimits_{X_1 }(c_1t) \times \cdots \times \mathop{\mathrm{cf}}\nolimits_{X_n }(c_nt).
\end{equation}

CF of a stochastic convolution defined by $Y = X_1 +\cdots + X_N$, where $X_j$ are i.i.d.~RVs with common $\mathop{\mathrm{cf}}\nolimits_{X}(t)$ and $N$ is an independent  discrete RV with $\mathop{\mathrm{cf}}\nolimits_{N}(t)$, is given by 
\begin{equation}\label{eq09}
	\mathop{\mathrm{cf}}\nolimits_{Y}(t) = \mathop{\mathrm{cf}}\nolimits_{N}\Big(-\mathtt{i} \log\left( \mathop{\mathrm{cf}}\nolimits_{X}(t)\right) \Big).
\end{equation}
Equivalent expressions based on using the moment generating functions (MGFs) and/or the probability generating functions (PGFs) have been derived elsewhere, see, e.g., equation (3.7) in \cite{kaas2008modern} and/or equation (7) in \cite{Shevchenko2010}.

Hence, CF of the ALD defined in (\ref{eq01}) and (\ref{eq02}), say $\mathop{\mathrm{cf}}\nolimits_{S}(t)$ or $\mathop{\mathrm{cf}}\nolimits_{F_S}(t)$,  is given by
\begin{equation}\label{eq10}
	\mathop{\mathrm{cf}}\nolimits_{S}(t) = \mathop{\mathrm{cf}}\nolimits_{N}\Big(-\mathtt{i} \log\left( \mathop{\mathrm{cf}}\nolimits_{X}(t)\right) \Big),
\end{equation}
where $\mathop{\mathrm{cf}}\nolimits_{N}(t)$ and $\mathop{\mathrm{cf}}\nolimits_{X}(t)$ denote the known CFs of frequency and severity distributions. 
Moreover, using equation (\ref{eq08}) CF  of the ALD of the (compound) portfolio, $L$, as defined in (\ref{eq03}), is 
\begin{equation}\label{eq11}
	\mathop{\mathrm{cf}}\nolimits_{L}(t) = \mathop{\mathrm{cf}}\nolimits_{S_1}(t) \times \cdots \times  \mathop{\mathrm{cf}}\nolimits_{S_p}(t),
\end{equation}
where 	
\begin{equation}\label{eq12}
	\mathop{\mathrm{cf}}\nolimits_{S_i}(t) = \mathop{\mathrm{cf}}\nolimits_{N_i}\Big(-\mathtt{i} \log\left( \mathop{\mathrm{cf}}\nolimits_{X_i}(t)\right) \Big),
\end{equation}
with $S_i$, $N_i$, and $X_i$, for  $i = 1,\dots,p$, as specified in (\ref{eq03}).

\renewcommand\arraystretch{1.75} 
\begin{table}[!t]
\vskip -5pt
\caption{Selected characteristic functions of the discrete (frequency) and the continuous (severity) univariate distributions frequently used as the components of the compound CFs of the ALD. CF of the generalized Pareto distribution is used for modeling/fitting the heavy tails of other distributions. Here, $U(a,b,z)$ denotes the confluent hypergeometric function of the second kind, defined for the complex argument $z\in\mathbb{C}$. %
\label{tab1}}
\small
\begin{center}
\begin{tabular}{  p{.4\columnwidth}  p{.5\columnwidth} }
\hline
Probability distribution & Characteristic function \cr \hline
Dirac   \newline $\mathop{Dirac}(\mu)$,  $\mu \in \mathbb{R}$ location &  
				$\mathop{\mathrm{cf}}\nolimits_{N}(t) =  \mathrm{e}^{\mathtt{i}t\mu}$ \cr			
Binomial \newline $\mathop{Bino}(n,p)$,  $n \in \mathbb{N}$ number of trials, $p\in(0,1)$ success probability&  
				$\mathop{\mathrm{cf}}\nolimits_{N}(t) =  \left(1-p+p \mathrm{e}^{\mathtt{i}t}\right)^n$ \cr
Negative Binomial \newline $\mathop{NeBi}(r,p)$, $r >0 $, $p\in(0,1)$ success probability&  
				$\mathop{\mathrm{cf}}\nolimits_{N}(t) =  p^r \left(1-(1-p) \mathrm{e}^{\mathtt{i}t}\right)^{-r}$ \cr
Poisson \newline $\mathop{Pois}(\lambda)$,  $\lambda > 0$ rate &  
				$\mathop{\mathrm{cf}}\nolimits_{N}(t) =  \exp\left( \lambda\left(\mathrm{e}^{\mathtt{i}t}-1\right)\right)$ \cr
\hline
Exponential \newline $\mathop{Expo}(\lambda)$, $\lambda > 0$ rate &  
				$\mathop{\mathrm{cf}}\nolimits_{X}(t) =  \frac{\lambda}{\lambda-\mathtt{i}t}$ \cr
Gamma \newline $\mathop{Gamm}(\alpha,\beta)$, 	$\alpha > 0$ shape, $\beta > 0$ rate, with scale $\sigma = \frac{1}{\beta}$ &  
				$\mathop{\mathrm{cf}}\nolimits_{X}(t) =  \left(1 - \frac{\mathtt{i}t}{\beta} \right)^{-\alpha}$ \cr
Log-normal \newline $\mathop{LogN}(\mu,\sigma^2)$, $\mu \in \mathbb{R}$ location, $\sigma > 0$ scale &  
				$\mathop{\mathrm{cf}}\nolimits_{X}(t)$ evaluated by combination of methods (\ref{eq05}) and/or (\ref{eq07}) with \newline
				$\mathop{\mathrm{pdf}}\nolimits_{X}(z) = \frac{1}{z\sigma\sqrt{2\pi}}\mathrm{e}^{-\frac{(\log(z)-\mu)^2}{2\sigma^2}}$ \cr
Log-logistic \newline $\mathop{LogL	}(\alpha,\beta)$, 	$\alpha > 0$ scale,\newline $\beta > 0$ shape, &  
				$\mathop{\mathrm{cf}}\nolimits_{X}(t)$ evaluated by combination of methods (\ref{eq05}) and/or (\ref{eq07}) with \newline
				$\mathop{\mathrm{pdf}}\nolimits_{X}(z) = \frac{\frac{\beta}{\alpha}\left( \frac{z}{\alpha}\right)^{\beta-1}}{\left(1+\left( \frac{z}{\alpha}\right)^\beta \right)^2}$ \cr
Pareto Type I \newline (European)\newline $\mathop{ParE}(\alpha,\sigma)$, $\alpha > 0$ shape, $\sigma > 0$ scale &  
				$\mathop{\mathrm{cf}}\nolimits_{X}(t) =  \alpha \mathrm{e}^{\mathtt{i}t \sigma}\mathop{U}\left(1,1-\alpha,-\mathtt{i}t \sigma \right)$ \newline 
				or $\mathop{\mathrm{cf}}\nolimits_{X}(t) = \mathrm{e}^{\mathtt{i}t\sigma}  \mathop{\mathrm{cf}_X^0}(t)$, where \newline
				$\mathop{\mathrm{cf}_X^0}(t)$ evaluated by combination of methods (\ref{eq05}) and/or (\ref{eq07}) with \newline
				$\mathop{\mathrm{pdf}}\nolimits_{X}(z) = \alpha \sigma^\alpha \left(\sigma+z\right)^{-(\alpha+1)}$ 
				\cr
Pareto Type II \newline (American, Lomax)\newline $\mathop{ParA}(\alpha,\sigma)$, $\alpha > 0$ shape, $\sigma > 0$ scale &  
				$\mathop{\mathrm{cf}}\nolimits_{X}(t) = \alpha \mathop{U}\left(1,1-\alpha,-\mathtt{i}t \sigma \right)$  \newline 
				or $\mathop{\mathrm{cf}}\nolimits_{X}(t)$ evaluated by combination of methods (\ref{eq05}) and/or (\ref{eq07}) with \newline
				$\mathop{\mathrm{pdf}}\nolimits_{X}(z) = \alpha \sigma^\alpha \left(\sigma+z\right)^{-(\alpha+1)}$ \cr
\hline
Generalized Pareto 
\newline $\mathop{GPD}(\xi,\sigma,\theta)$, here $\xi \geq 0$ shape, $\sigma > 0$ scale, $\theta \geq 0$ threshold &  
				$\mathop{\mathrm{cf}}\nolimits_{X}(t) = \mathrm{e}^{\mathtt{i}t\theta}  \mathop{\mathrm{cf}_X^0}(t)$, where \newline
				$\mathop{\mathrm{cf}_X^0}(t)$ evaluated by combination of methods (\ref{eq05}) and/or (\ref{eq07}) with \newline
				$\mathop{\mathrm{pdf}}\nolimits_{X}(z) = \frac{1}{\sigma}\left(1+\xi\frac{z}{\sigma} \right)^{-(\frac{1}{\xi}+1)}$ \cr
\hline
\end{tabular}
\end{center}
\vskip -15pt
\end{table}
\renewcommand\arraystretch{1.0} 

Finally, CF of a weighted mixture distribution, defined by $F_Y=\sum_{j=1}^n  w_j F_{X_j}$ with $\sum_{j=1}^n  w_j = 1$, is 
\begin{equation}\label{eq13}
	\mathop{\mathrm{cf}}\nolimits_{Y}(t) = \sum_{j=1}^n  w_j\mathop{\mathrm{cf}}\nolimits_{X_j }(t),
\end{equation}
where $\mathop{\mathrm{cf}}\nolimits_{Y}(t)$ denotes the CF of the distribution $F_Y$ and $\mathop{\mathrm{cf}}\nolimits_{X_j }(t)$ denotes the CF of the distribution $F_{X_j}$, for $j = 1,\dots,n$.

For illustration, Table~\ref{tab1} presents selected CFs of the discrete and the continuous univariate distributions frequently used as the building blocks of the compound CFs of the aggregate loss distributions, as defined in (\ref{eq10}), (\ref{eq11}) and (\ref{eq12}).

\subsection{Empirical CF}

Let $X_1, \dots, X_n$ are i.i.d.~random variables with the common distribution function $F_X$. 
The empirical distribution based on the random sample $X_1, \dots, X_n$ is a mixture distribution of equally weighted degenerate Dirac distributions, concentrated at $X_1, \dots, X_n$. 

Hence, the observed empirical characteristic function (ECF) is equally weighted mixture of the characteristic functions of the Dirac random variables concentrated at the observed values $x_j$ of $X_j$, i.e.~mixture of CFs given by $\mathop{\mathrm{cf}}\nolimits_{x_j}(t) = \mathrm{e}^{\mathtt{i}tx_j}$, 
\begin{equation}\label{eq14}
	\mathop{\mathrm{cf}}\nolimits_{\hat{F}_X}(t) = \frac{1}{n}\sum_{j=1}^n  \mathrm{e}^{\mathtt{i}tx_j}. 
\end{equation}

Let $\hat{F}_N$ denotes the empirical CDF (ECDF) of the observed historic numbers (frequency) of claims $n_1,\dots,n_J$, in each of $J$ historic years, with its empirical CF given by
\begin{equation}\label{eq15}
	\mathop{\mathrm{cf}}\nolimits_{\hat{F}_N}(t) = \frac{1}{J}\sum_{j=1}^J  \mathrm{e}^{\mathtt{i}tn_j}. 
\end{equation}
Further, let $\hat{F}_X$ denotes the ECDF based on $K$ observed historic values (severity) of claims $x_1,\dots,x_K$, with its ECF given by
\begin{equation}\label{eq16}
	\mathop{\mathrm{cf}}\nolimits_{\hat{F}_X}(t) = \frac{1}{K}\sum_{k=1}^K  \mathrm{e}^{\mathtt{i}tx_k}.  
\end{equation}
Then, in analogy with (\ref{eq10}), the compound empirical CF, say $\mathop{\mathrm{cf}}\nolimits_{\hat{F}_S}(t)$, of the collective risk $S$ distribution is 
\begin{equation}\label{eq17}
	\mathop{\mathrm{cf}}\nolimits_{\hat{F}_S}(t) = \mathop{\mathrm{cf}}\nolimits_{\hat{F}_N}\Big(-\mathtt{i} \log\left( \mathop{\mathrm{cf}}\nolimits_{\hat{F}_X}(t)\right) \Big) 
	=\frac{1}{J} \sum_{j=1}^J \left(\frac{1}{K} \sum_{k=1}^K  \mathrm{e}^{\mathtt{i}tx_k}  \right)^{n_j}.
\end{equation}

Similarly, and in analogy with (\ref{eq11})-(\ref{eq12}), we can also derive the empirical CF of the ALD of the (compound) portfolio, $L$, as defined in (\ref{eq03}), say $\mathop{\mathrm{cf}}\nolimits_{\hat{F}_L}(t)$.

Combination of the ECDF $\hat{F}_X$ and the fitted generalized Pareto CDF is frequently used for modeling the heavy tailed (severity) distributions, based on the observed data, see, e.g., \cite{mcneil1997estimating} and \cite{mcneil1997peaks}. Using equation (\ref{eq13}) the CF of such distribution, say $\mathop{\mathrm{cf}}\nolimits_{\widehat{F_X}}(t)$,  can be expressed as a weighted mixture of the empirical CF and the generalized Pareto CF, 
\begin{equation}\label{eq18}
	\mathop{\mathrm{cf}}\nolimits_{\widehat{F_X}}(t) = p  \mathop{\mathrm{cf}}\nolimits_{\hat{F}_{X_L}}(t) + (1-p)  \mathop{\mathrm{cf}}\nolimits_{\mathop{GPD}}(t),
\end{equation}
where $p\in(0,1)$ is chosen probability level specifying the tail part of the distribution, typically with $p=0.9$ or greater, $\mathop{\mathrm{cf}}\nolimits_{\hat{F}_{X_L}}(t)$ is the empirical CF based on the lower $p$-part of the observed values ($x_k \leq \theta$, where $\theta$ is the threshold selected as the $p$-quantile of the distribution), and $\mathop{\mathrm{cf}}\nolimits_{\mathop{GPD}}(t)$ is CF of the fitted generalized Pareto distribution $\mathop{GPD}(\xi,\sigma,\theta)$, with the parameters $\xi$ and $\sigma$ estimated (e.g., by the maximum likelihood estimation method) from the observed values $x_k-\theta \geq 0$, $k = 1,\dots,K$.

Finally, notice that the ALD defined by the CF (\ref{eq10}) or (\ref{eq11})-(\ref{eq12}) is a discrete distribution, if all of the  severity component distributions ($S$ or $S_m$ for all $m=1,\dots,M$) are discrete distributions (e.g., based on their empirical CFs). Otherwise, the ALD distribution is a continuous distribution (although with possibly highly erratic shape). In particular, the aggregate loss distribution defined by the empirical CF (\ref{eq17}) is in principal a discrete one with cumulative distribution function being a step-function (similarly as is the empirical CDF).

The standard inversion theorems, including the Gil-Pelaez inversion formulae introduced below, are based on the assumption that the PDF exists (i.e. assuming the absolutely continuous distribution) and that the characteristic function is absolutely integrable over $(-\infty,\infty)$. 
In Section~\ref{sec03} we present methods and algorithms that are based on this theoretical assumption, however, for most practical purposes, their numerical implementation is typically also well suited (as an approximate method) for evaluation of the ALD CDFs defined by the empirical CFs. 

In any case, smoothing the compound loss empirical ALD, $L$, is still possible by using the appropriately smoothed CF, obtained by convolution of the empirical distribution with suitable continuous distribution, as
\begin{equation}\label{eq19}
\mathop{\mathrm{cf}}\nolimits_{\tilde{F}_L}(t) = \mathop{\mathrm{cf}}\nolimits_{\hat{F}_L}(t) \times  \mathop{\mathrm{cf}}\nolimits_{Z}(t),
\end{equation}
where $\mathop{\mathrm{cf}}\nolimits_{\hat{F}_L}(t)$ is the empirical or otherwise corrupted CF, and $\mathop{\mathrm{cf}}\nolimits_{Z}(t)$ represents CF of a suitable \emph{smoothing} continuous distribution, e.g., zero-mean Gaussian distribution with its standard deviation $\sigma$ proportional to the selected bandwidth of the smoothing kernel, $\mathop{\mathrm{cf}}\nolimits_{Z}(t) =  \mathrm{e}^{-\sigma^2t^2/2}$.

\section{The Gil-Pelaez inversion formulae}\label{sec03}

\subsection{Numerical evaluation}

Computing the (inverse) Fourier transform numerically is a well-known problem, frequently connected with computing integrals of highly oscillatory (complex) functions. It was studied for a long time in general, but also with focus on specific applications, see, e.g., \cite{asheim2013complex,levin1996fast,milovanovic1998numerical,sidi1982numerical,sidi1988user,sidi2012user}, 
 to show just a few. In particular, the methods suggested for inverting the characteristic function for obtaining the probability distribution function include 
\cite{abate1992fourier,shephard1991characteristic,waller1995obtaining,zielinski2001high,strawderman2004computing}, and \cite{feng2013inverting}. 

Here we shall assume that the considered ALD characteristic function, say $\mathop{\mathrm{cf}}\nolimits_{L}(t)$ or $\mathop{\mathrm{cf}}\nolimits_{F_L}(t)$, which is associated with the  distribution of the specific aggregate loss $L\sim F_L$, is known and can be easily evaluated for arbitrary $t\in \mathbb{R}$.

\cite{GilPelaez1951} derived the inversion formulae of the absolutely integrable CFs over $(-\infty,\infty)$, suitable for numerical evaluation of the PDF and/or the CDF, which require integration of a real-valued functions only, for more details see \cite{shephard1991characteristic}. In particular,  PDF of the absolutely continuous distribution (assuming that it exists), with characteristic function $\mathop{\mathrm{cf}}\nolimits_{L}(t)$, is given by
\begin{equation}\label{eq20}
\mathop{\mathrm{pdf}}\nolimits_{L}(\ell) = \frac{1}{\pi}\int_0^\infty \Re\left(\mathrm{e}^{-\mathtt{i}t\ell}\mathop{\mathrm{cf}}\nolimits_{L}(t)  \right)\,dt,
\end{equation}
and further, if $\ell$ is a continuity point of the cumulative distribution function of $L$, defined by $\mathop{\mathrm{cdf}}\nolimits_{L}(\ell) = \Pr\{L\leq \ell\}$, then the CDF is given by
\begin{equation}\label{eq21}
\mathop{\mathrm{cdf}}\nolimits_{L}(\ell)  =  \frac{1}{2}-\frac{1}{\pi}\int_0^\infty \Im\left(\frac{\mathrm{e}^{-\mathtt{i}t\ell}\mathop{\mathrm{cf}}\nolimits_{L}(t) }{t} \right)\,dt.
\end{equation}
By $\Re(f(t))$ and $\Im(f(t))$ we denote the real and imaginary part of the complex function $f(t)$, respectively.

In statistics, numerical inversion of the characteristic function based on (\ref{eq20}) and (\ref{eq21}) was successfully implemented in another context for evaluation of the distribution function of a linear combination of independent chi-squared RVs in \cite{Imhof1961} and \cite{Davies1980}. Further, the Gil-Pelaez's method was used to compute the distribution of a linear combination of independent Student's $t$ random variables and also for the distribution of a linear combination of independent inverted gamma random variables, see  \cite{Witkovsky2001a}, \cite{Witkovsky2001b}, and also \cite{Witkovsky2015}.

In general, the integrals in (\ref{eq20}) and (\ref{eq21}) can be computed by a number of numerical quadrature methods. In some the integral is subdivided into subintervals between consecutive zeroes of the integrand, integrated over them using, for example, Gaussian quadrature, and the summation of the obtained alternating series is accelerated by known methods, see, e.g., \cite{cohen2000convergence}. 

Frequently, however, the integral (\ref{eq20}) and in particular (\ref{eq21}) can be efficiently approximated by a trapezoidal quadrature, i.e.
\begin{equation}\label{eq22}
	\mathop{\mathrm{pdf}}\nolimits_{L}(\ell) \approx \frac{\delta}{\pi}\sum_{j=0}^N w_j \Re\left(\mathrm{e}^{-\mathtt{i}t_j\ell}\mathop{\mathrm{cf}}\nolimits_{L}(t_j)  \right),
\end{equation}
and/or
\begin{equation}\label{eq23}
\mathop{\mathrm{cdf}}\nolimits_{L}(\ell) \approx	\frac{1}{2}-\frac{\delta}{\pi} \sum_{j=0}^N w_j  \Im\left(\frac{\mathrm{e}^{-\mathtt{i}t_j\ell}\mathop{\mathrm{cf}}\nolimits_{L}(t_j) }{t_j} \right),
\end{equation}
where $N$ is sufficiently large integer, say $N = 2^{10}$, $w_j$ are the appropriate trapezoidal quadrature weights (i.e.~$w_0= w_{N} = \frac{1}{2}$, and $w_j=1$ for $j=1,\dots,N-1$), and $t_j = j\delta$ for $j=0,\dots,N$ are the equidistant nodes (with their mutual distance $\delta$) from the interval $[0,T]$, for sufficiently large $T$ (i.e.~such $T$ that the integrand function 
$\Re\left(\mathrm{e}^{-\mathtt{i}t_j\ell}\mathop{\mathrm{cf}}\nolimits_{L}(t)  \right)$ and/or $\Im\left({\mathrm{e}^{-\mathtt{i}t_j\ell}\mathop{\mathrm{cf}}\nolimits_{L}(t) }/{t} \right)$ 
is sufficiently small for all $t> T$). 

Particular selection of the values $N$ and $T$ influences the total approximation error, i.e.~combination of the truncation error and the integration error. The trade-off between them strongly depends on $\mathop{\mathrm{cf}}\nolimits_{L}$. 

If the optimum values of $N$ and $T$ are unknown, we suggest, as  a simple rule of thumb, to start with the application of the following \emph{six-sigma-rule}. 

For that, set $\delta = {2\pi}/{(B-A)}$, where the interval $(A,B) = \mathop{E}(L)\mp k\sqrt{\mathop{\mathrm{Var}}(L)}$ with $k=6$ (or other more suitable value of the multiplication coefficient $k$) specifies the substantial part of the distribution support of the random variable $L$, and then set $N$ and $T = N \delta$ such that the absolute value of the integrand function is sufficiently small for all $t> T$, say $\left|\Im\left({\mathrm{e}^{-\mathtt{i}t_j\ell}\mathop{\mathrm{cf}}\nolimits_{L}(t) }/{t} \right)\right| \leq \left|\mathop{\mathrm{cf}}\nolimits_{L}(t) /t \right|
< \varepsilon$ with, e.g., $\varepsilon = 10^{-12}$.

Further, for computing the first term in (\ref{eq23}), we can use the result from \cite{Witkovsky2001a}: If the mean (expectation) of $L$ exists, then 
\begin{equation}\label{eq24}
	\lim_{t\rightarrow 0} \Im\left(\frac{\mathrm{e}^{-\mathtt{i}t\ell}\mathop{\mathrm{cf}}\nolimits_{L}(t) }{t} \right) = \mathop{E}(L)-\ell. 
\end{equation}

The required location and dispersion parameters, the expectation $\mathop{E}(L)$ and the variance $\mathop{\mathrm{Var}}(L)$, can be evaluated either analytically, from the moments of the distribution (i.e.~the  expectation and the variance of $L$, if they exist and are known), or approximately, by numerical differentiation of the (known) $\mathop{\mathrm{cf}}\nolimits_{L}(t)$. 
In particular, for any small $h>0$, e.g., $h = 10^{-4}$, we get
\begin{equation}\label{eq25}
		\mathop{E}(L) \approx \frac{1}{h}
	\left(\frac{8}{5}\Im\Big(\mathop{\mathrm{cf}}\nolimits_{L}(h)\Big) - \frac{2}{5}\Im\Big(\mathop{\mathrm{cf}}\nolimits_{L}(2h)\Big)  
	+ \frac{8}{105}\Im\Big(\mathop{\mathrm{cf}}\nolimits_{L}(3h)\Big) - \frac{2}{280}\Im\Big(\mathop{\mathrm{cf}}\nolimits_{L}(4h)\Big) 
	\right),
\end{equation}
and, 
\begin{equation}\label{eq26}
	\mathop{\mathrm{Var}}(L) \approx \mathop{E}\left(L^2\right)-\Big(\mathop{E}(L)\Big)^2, 
\end{equation}
where
\begin{equation}\label{eq27}
\mathop{E}\left(L^2\right) \approx \frac{1}{h^2} 
	\left(\frac{205}{72} -
	\frac{16}{5}\Re\Big(\mathop{\mathrm{cf}}\nolimits_{L}(h)\Big) + \frac{2}{5}\Re\Big(\mathop{\mathrm{cf}}\nolimits_{L}(2h)\Big) 
  - \frac{16}{315}\Re\Big(\mathop{\mathrm{cf}}\nolimits_{L}(3h)\Big) + \frac{2}{560}\Re\Big(\mathop{\mathrm{cf}}\nolimits_{L}(4h)\Big) \right).
\end{equation}

Such approximation, based on numerical differentiation of the characteristic function, serve as a reasonably good approximation of the required location and scale parameters also in situations when the theoretical moments (expectation and variance) formally do not exist.

Note that the presented quadrature is efficient fot computing $\mathop{\mathrm{pdf}}\nolimits_{L}(\ell)$ and $\mathop{\mathrm{cdf}}\nolimits_{L}(\ell)$ for any $\ell \in (A,B)$ as it requires only one evaluation of $\mathop{\mathrm{cf}}\nolimits_{L}(t_j)$ for $j = 0,\dots,N$.

Finally, the \emph{quantile function} (QF) used for computing the VaRs can be evaluated either by simple interpolation from values calculated by (\ref{eq23}), or (for sufficiently smooth continuous distributions) by the iterative Newton-Raphson scheme. It requires repeated evaluations of the PDF/CDF (\ref{eq22})-(\ref{eq23}). In particular, for fixed probability level $p\in(0,1)$, the $p$-quantile of the (continuous) distribution of $L$, say $q=\mathop{\mathrm{qf}}\nolimits_{L}(p)$, is given as a solution (fixed point) of the following iterative scheme, 
\begin{equation}\label{eq28}
\mathop{\mathrm{qf}}\nolimits_{L}^{(k+1)}(p) = \mathop{\mathrm{qf}}\nolimits_{L}^{(k)}(p)  - \frac{\mathop{\mathrm{cdf}}\nolimits_{L} \left(\mathop{\mathrm{qf}}\nolimits_{L}^{(k)}(p)\right)-p}{\mathop{\mathrm{pdf}}\nolimits_{L}\left(\mathop{\mathrm{qf}}\nolimits_{L}^{(k)}(p)\right)},	
\end{equation}
where $k=0,1,\dots$, and the starting value $\mathrm{qf}_{L}^{(0)}(p)$ is set as, e.g., $\mathrm{qf}_{L}^{(0)}(p) = \mathop{E}(L)$, given by (\ref{eq25}).
 
\subsection{Software implementation}

We have implemented the above mentioned methods and algorithms into the {MATLAB} \emph{characteristic functions toolbox} (\texttt{CF Toolbox}).  It is a set of algorithms for computing and combining the characteristic functions and further for computing the PDF, CDF, and QF, by numerical inversion of the associated CF. The toolbox is available 
from the authors at the web page: \url{https://goo.gl/gBfdwY}.

The \texttt{CF Toolbox} includes also the easy to use application,  the \emph{collective risk model tool} (\texttt{CRM Tool}) . The \texttt{CRM Tool} is a fast and for most practical situations reasonably precise calculator of the aggregate claim/loss distribution and the associated value at risk, specified and computed by numerical inversion of its characteristic function. 

The algorithms used in the \texttt{CF Toolbox} are based on trapezoidal rule for computing the integrals defined by the Gil-Pelaez formulae, or by using the FFT algorithm for computing the Fourier transform integrals. As already mentioned, in more complicated situations or if the highest numerical precision is required, a more advanced quadrature methods combined with accelerated computing of limits of series with alternating signs are typically required. For more details on possible alternative approaches and {MATLAB} implementation see \cite{Duby2017}.

\section{Real data example: The Danish fire losses data}\label{sec04}

\begin{table*}
	\centering
		\begin{tabular}{lrrrrrrrrrrr}
\hline\cr
Year      &	1980	& 1981	& 1982  & 1983  &	1984  &	1985  &	1986  &	1987 &	1988	& 1989  &	1990 \cr
Number of claims &	166	  & 170	  & 181	  & 153   &	163   &	207	  & 238   &	226  &	210	  & 235	  & 218 \\[3mm]
\hline
		\end{tabular}
	\caption{The Danish fire losses data: Number of claims (the empirical frequency distribution) observed during the period 1980-1990.}
	\label{Tab01}
\end{table*}

\begin{figure*}[t]
\begin{center}
\includegraphics[width = 0.95\textwidth]{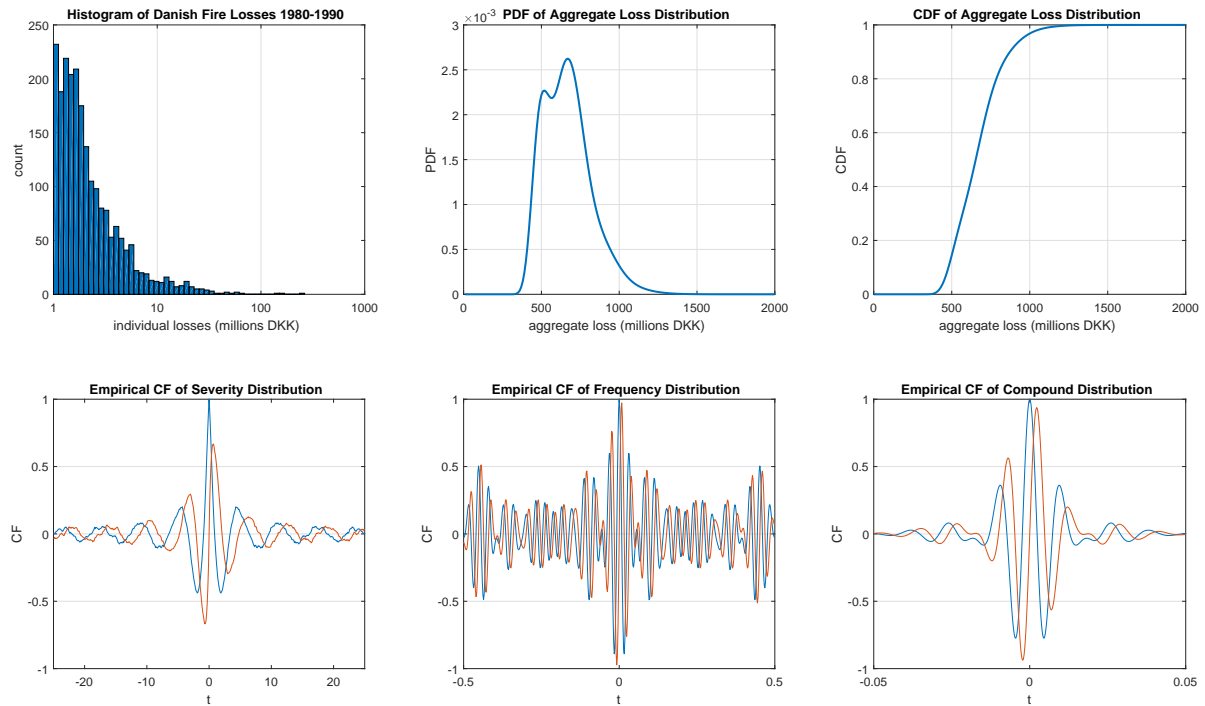}
\end{center}
\caption{The Danish fire losses data: (i) Histogram of the individual losses in millions DKK observed during the period 1980-1990 (the severity distribution), presented in the logarithmic scale. (ii) PDF of the aggregate loss distribution derived by numerical inversion from the empirical compound CF. (iii) CPDF of the aggregate loss distribution derived by numerical inversion from the empirical compound CF. (iv) Real (blue) and imaginary part (red) of the empirical characteristic function of the severity distribution. (v)  The empirical characteristic function of the frequency distribution. (vi) The compound empirical characteristic function of the aggregate loss distribution.}\label{Fig01}
\end{figure*}

For illustration purposes, here we present the analysis of a well known insurance dataset frequently used for comparison of methods: the data on major Danish fire insurance losses, see \cite{Eling2012} and references therein. The dataset is comprised of Danish fire losses originally analyzed in \cite{mcneil1997estimating} and \cite{resnick1997discussion}. The data represents fire losses in million Danish Krones (DKK) and was collected by a Danish reinsurance company. The dataset contains individual losses above 1 million DKK, a total of $2167$ individual losses, covering the period from January 3, 1980 to December 31, 1990. It is adjusted for inflation to reflect 1985 values. The dataset can be found in the R packages \texttt{fEcofin} and \texttt{fExtremes} and is also included in the {MATLAB} \texttt{CF Toolbox}. 

The empirical frequency distribution, the number of claims per year during the period 1980-1990, is given in the Table~\ref{Tab01} (mean value of $197$ claims per year). However, it is clearly visible that during the period 1980-1985 the number of claims was lower (mean value $166.6$) than the number of claims during the period 1986-1990 (mean value $222.3$). This suggest possible mixture of different random mechanisms generating the number of claims, which are difficult to model by a standard discrete distribution.

The empirical severity distribution, based on losses of $2167$ individual claims greater than 1 million DKK observed during this period is presented as a histogram (in logarithmic scale) in the upper left panel of Figure~\ref{Fig01}. The descriptive statistics show that the distribution of the individual fire losses are significantly skewed to the right and exhibit high kurtosis, with the observed mean of $3.39$ and standard deviation of $8.51$ (millions DKK), skewness $18.74$ and kurtosis of $485.65$. This suggest to consider a heavy tail distribution as a model of the severity distribution.

The first modeling approach for deriving the ALD is based on a purely nonparametric approach for deriving the aggregate loss distribution from the compound empirical characteristic functions (\ref{eq15}), (\ref{eq16}), and (\ref{eq17}) by numerical inversion (\ref{eq22})--(\ref{eq23}). 

With \texttt{CF Toolbox} the evaluation of the aggregate loss distribution (PDF/CDF) specified by its CF as well as of the required VaRs is a simple task, which can be formulated by several lines of MATLAB code:

{\small
\begin{verbatim}
% Danish fire losses data:
load('DanishFireData.mat')

% Empirical characteristic functions:
cfN  = @(t) cfE_Empirical(t,Frequency);
cfX  = @(t) cfE_Empirical(t,Severity);
cf   = @(t) cfN(-1i*log(cfX(t)));

% Parameters/options:
prob = [0.9 0.99 0.999];
loss = linspace(0,2000,201)';
options.isCompound = true;

% Numerical inversion of CF by cf2DistGP:
result = cf2DistGP(cf,loss,prob,options); 
\end{verbatim}}

The outcome of the calculation is a MATLAB structure array (\texttt{result}) with specified fields and values which contain the values of PDF and CDF evaluated at required values (specified by the variable \texttt{loss}), as well as the values of the VaRs evaluated at the required probabilities (specified by the variable \texttt{prob}). In particular, the calculated values at risk (VaRs) evaluated for the probabilities $0.9$, $0.99$, and $0.999$ are: $872.9$, $1112.8$, and $1319.6$ (in millions DKK).

The second modeling approach for deriving ALD is based on a semi-parametric approach, by incorporating the generalized Pareto distribution fit of the severity distribution heavy tails. Here, the compound CF is 
\begin{equation}\label{eq29}
\mathop{\mathrm{cf}}\nolimits_{\widehat{F_S}}(t) = \mathop{\mathrm{cf}}\nolimits_{\hat{F}_N}\Big(-\mathtt{i} \log\left( \mathop{\mathrm{cf}}\nolimits_{\widehat{F_X}}(t)\right) \Big), 
\end{equation}
where $\mathop{\mathrm{cf}}\nolimits_{\hat{F}_N}$ is given by (\ref{eq15}) and $\mathop{\mathrm{cf}}\nolimits_{\widehat{F_X}}$ is a weighted mixture of the empirical CF and the fitted generalized Pareto CF, given by (\ref{eq18}). Then, the ALD is derived by numerical inversion (\ref{eq22})--(\ref{eq23}) from the compound CF (\ref{eq29}). 

Choice of the optimum threshold $\theta$, which divides the observed losses (severity data) into the head (main body) area and the tail area used for fitting the generalized Pareto tail distribution, is the hard part of this modeling approach, which is discussed in more details elsewhere, see, e.g., \cite{mcneil1997estimating} where the threshold values between $10$ and $20$ have been considered. For simplicity, however, here we consider the threshold derived as a $p$-quantile of the empirical severity distribution, specified by the probability value $p=0.95$. For given fire losses data we get the estimate $\hat{\theta} = 10.0203$ (millions DKK).

Then the fitted generalized Pareto distribution $\mathop{GPD}\left(\hat{\xi},\hat{\sigma},\hat{\theta}\right)$ has the parameters (estimated by the maximum likelihood estimation method from the observed losses greater than $\hat{\theta}= 10.0203$): $\hat{\xi} = 0.4890$ and $\hat{\sigma} = 7.1082$. 

Based on that, we can construct the severity distribution characteristic function $\mathop{\mathrm{cf}}\nolimits_{\widehat{F_X}}$, defined as a mixture of the empirical CF and the fitted generalized Pareto CF, as given in (\ref{eq18}), and with the empirical frequency CF $\mathop{\mathrm{cf}}\nolimits_{\hat{F}_N}$ also the compound characteristic function $\mathop{\mathrm{cf}}\nolimits_{\widehat{F_S}}$.

As before, evaluation of the aggregate loss distribution specified by this compound CF can be formulated by several lines of MATLAB code:

{\small
\begin{verbatim}
% Set the threshold parameter theta
p     = 0.95;
theta = quantile(Severity,p);

% Fit the GP (Generalized Pareto) distribution
GPfit = paretotails(Severity,0,p);
Pars  = GPfit.UpperParameters;
xi    = Pars(1);
sigma = Pars(2);

% CF of the fitted tail GP distribution 
pdfGP = @(x) gppdf(x,xi,sigma);
cfGP  = @(t) cfX_PDF(t,pdfGP) .* exp(1i*t*theta);

% CF of the mixture severity distribution
XL    = Severity(Severity <= theta);
cfXL  = @(t) cfE_Empirical(t,XL);
cfX   = @(t) p * cfXL(t) + (1-p) * cfGP(t);

% Empirical CF of the frequency distribution
cfN   = @(t) cfE_Empirical(t,Frequency);

% Compound CF of the aggregate loss distribution
cf    = @(t) cfN(-1i*log(cfX(t)));

% Parameters
prob = [0.9 0.99 0.999];
loss = linspace(0,2500,201)';

% Options
clear options
options.N = 2^16;
options.SixSigmaRule = 15;
options.isCompound = true;

% Numerical inversion of CF by cf2DistGP
result = cf2DistGP(cf,loss,prob,options); 
\end{verbatim}}

The calculated VaRs (values at risk) evaluated for the probabilities $0.9$, $0.99$, and $0.999$ are: $847.96$, $1156.8$, and $2063.3$ (in millions DKK). 

These VaRs (especially the higher quantiles) are different if compared with the VaRs estimated from the purely nonparametric approach based on inverting the empirical compound CF. In general, as pointed out in \cite{mcneil1997estimating}: \emph{Every dataset is unique and the data analyst must consider what the data mean at every step. The process cannot and should not be fully automated}. 

\section{Conclusions}\label{sec05}
We propose numerical inversion methods for derivation of the aggregate loss distribution from its characteristic function, derived as compound characteristic function of the frequency CF and the severity CF. 

In particular, in this paper we emphasize the nonparametric approach based on using the empirical characteristic functions of the frequency distribution and the severity distribution of the claims in the actuarial risk applications. 

As was illustrated, this can be generalized into a more complex semi-parametric modeling approach by incorporating the generalized Pareto distribution fit of the severity distribution heavy tails, and/or by considering the weighted mixture of the parametric CFs (used to model the expert knowledge) and the empirical CFs (used to incorporate the knowledge based on the historical data).

The presented numerical inversion method is based on combination of the Gil-Pelaez inversion formulae and the simple trapezoidal rule used for numerical integration. The methods and algorithms are incorporated in the {MATLAB} \emph{characteristic functions toolbox} (\texttt{CF Toolbox}), which is available 
at the web page \url{https://goo.gl/gBfdwY}.

The applicability of the suggested approach was illustrated by analysis of a well know insurance dataset, the Danish fire loss data. As it was emphasized in \cite{mcneil1997estimating}, such inference is very sensitive to the choice of the threshold and also to the largest observed losses, and thus, the process cannot and should not be fully automated. There is a role for stress scenarios in such loss severity analyses, whereby historical loss data are enriched by hypothetical losses to investigate the consequences of unobserved, adverse events.

The suggested methods and algorithms could serve very well for this purpose. 

\section*{Acknowledgement}
The work was supported by the Slovak Research and Development Agency, project APVV-15-0295, and by the Scientific Grant Agency VEGA of the Ministry of Education of the Slovak Republic and the Slovak Academy of Sciences, by the projects VEGA 2/0047/15 and VEGA 2/0011/16.



\end{document}